\documentclass{article}

\usepackage{arxiv}

\usepackage[utf8]{inputenc}
\usepackage[T1]{fontenc}
\usepackage{xurl}
\usepackage{booktabs}
\usepackage{amsfonts}
\usepackage{nicefrac}
\usepackage{microtype}
\usepackage{graphicx}
\usepackage{natbib}
\usepackage{doi}
\usepackage{amsmath,amssymb}
\usepackage{placeins}
\usepackage{hyperref}
\usepackage{float}

\usepackage{silence}
\title{MATCH: Multi-stage Article Tagging to Concept Hierarchies}

\author{
{\normalfont Daniele Raimondi}\\
\texttt{daniele.raimondi@mdpi.com}
\and
{\normalfont Feichi Lu}\\
\texttt{feichi.lu@mdpi.com}
\and
{\normalfont Oliver Grun}\\
\texttt{oliver.grun@mdpi.com}
\and
{\normalfont Mariia Eremina}\\
\texttt{mariia.eremina@mdpi.com}
\and
{\normalfont Andrea Perlato\thanks{Corresponding author.}}\\
\texttt{andrea.perlato@mdpi.com}
}

\date{MDPI Data Intelligence Team \\
MDPI AG \\
Grosspeteranlage 5 \\
CH-4052 Basel, Switzerland
}

\renewcommand{\headeright}{}

\hypersetup{
pdftitle={MATCH: Multi-stage Article Tagging to Concept Hierarchies},
pdfsubject={cs.DL, cs.IR},
pdfauthor={Daniele Raimondi, Feichi Lu, Oliver Grun, Mariia Eremina, Andrea Perlato},
pdfkeywords={Article tagging, Knowledge organization, Subject indexing, Text embeddings, Large language models, OpenAlex},
}

\begin{document}
\maketitle

\begin{abstract}
The increasing specialization of scientific research calls for fine-grained representations of publication content that can be applied at scale. SCALE extends the OpenAlex taxonomy with approximately 114,000 Scientific Concepts derived from author keywords, but assigning these Concepts to individual publications remains challenging. We introduce MATCH (Multi-stage Article Tagging to Concept Hierarchies), a framework that assigns SCALE Concepts using article titles, abstracts, and author keywords. MATCH combines multi-view embedding retrieval, deterministic taxonomy-aware filtering, and language-model reranking to select relevant Concepts while preserving their hierarchical paths. Its modular design supports both API-based and self-hosted inference. In a blind evaluation involving 22 domain experts across 11 fields and 110 articles, the self-hosted and API-based configurations achieved Precision of 81.5\% and 83.3\%, respectively, and Precision@5 of approximately 89\% for both, with no statistically significant difference detected. These findings support the practical viability of the evaluated self-hosted configuration for large-scale Concept assignment without per-token API charges. Deployed on over two million MDPI publications, MATCH assigns a median of five Concepts per article, covers approximately 94\% of the SCALE Concepts, and supports daily indexing of newly available publications. By operationalizing the SCALE taxonomy at publication level, MATCH provides a foundation for fine-grained scientometric analysis and monitoring of scientific developments.
\end{abstract}


\clearpage

\section{Introduction}
\label{sec:introduction}

As the scholarly literature becomes increasingly specialized, describing the focus of individual articles accurately becomes more important.
Broad disciplinary categories, particularly those used at the journal level, may be too general to capture the specific topic of a paper, especially in multidisciplinary research \citep{rivest2021}.
Article-level classification therefore provides a more precise view of what each publication is about.
Publication-level systems, built from citation relations or by reassigning articles to disciplinary categories, address the limitations of journal-level classification \citep{waltman2012,milojevic2020}, and algorithmic classifications have been studied at progressively finer granularity \citep{sjogarde2018}.
At this level of detail, the task is not just to extract terms from the text.
It also requires assigning descriptors from an organized set of terms that reflect the article's subject \citep{golub2021}.
Representing an article through several specific concepts describes its scientific focus more precisely than one broad category, and a shared concept vocabulary reduces the variability introduced by author keywords.
This approach can make publications easier to compare and provide a stronger basis for scientometric analysis and literature exploration.

To support this type of representation, SCALE extended the OpenAlex taxonomy by adding about 114,000 Scientific Concepts below the existing Topics \citep{raimondi2026scale}.
One of the main goals of this layer is to enable fine-grained classification at the article level.
However, a taxonomy only defines which Concepts exist and how they are organized. To support scientometric analysis and literature exploration, it must also be known which publications belong to which Concepts, and assigning them is a separate methodological problem.
Selecting only a few labels from such a large space is a complex hierarchical multi-label problem.
This becomes even more difficult when many categories are highly specialized and rare \citep{sadat2022,babbar2019}.
In this paper, we introduce MATCH, a framework for linking each paper to a limited number of relevant Concepts.

MATCH addresses this problem through a multi-stage procedure.
The method uses the title, abstract, and author keywords to identify a broad set of potentially relevant Concepts.
Deterministic taxonomy-based filters then reduce the set of candidates and remove inconsistent or redundant assignments.
Finally, a language model selects the most relevant Concepts and ranks them.
We apply MATCH to more than two million articles and evaluate its assignment quality with 22 domain experts across 11 scientific fields, using both self-hosted and API-based language models to assess the framework's flexibility. MATCH is deployed in production and supports daily updates to assign Concepts to newly available articles.
Through this large-scale application and expert evaluation, we investigate whether publications can be reliably assigned to a taxonomy of approximately 114{,}000 fine-grained Concepts at the scale of millions of articles.

\section{Related Work}
\label{sec:relatedwork}
Previous research relevant to this study addresses three areas: assignment of controlled-vocabulary terms to publications, hierarchical and extreme multi-label classification (including retrieve-then-rerank approaches), and the step from taxonomy construction to article assignment.

\subsection{Subject Indexing and Article-Level Classification}
\label{subsec:subject-indexing}
Subject indexing aims to represent the content of a document through terms or categories from a knowledge organization framework. In scientometrics, assigning subject categories directly to individual publications can provide a more precise representation than journal-based approaches \citep{waltman2012,milojevic2020,rivest2021}. A key distinction is between derived indexing and assigned indexing. In derived indexing, terms are taken directly from the text. In assigned indexing, descriptors are selected from a controlled vocabulary, thesaurus, or classification system \citep{golub2019automatic,golub2021}. For MATCH, assigned indexing is especially relevant because the task is to link each article to labels that are already defined within a common structure.

Subject indexing with a fine-grained controlled vocabulary has a strong precedent in biomedicine, where PubMed articles are indexed with MeSH terms. Automating this process has been studied extensively, with supervised systems trained on large collections of manually indexed articles \citep{mao2017meshnow,you2021bertmesh}, and the U.S. National Library of Medicine has moved MEDLINE indexing to an automated process \citep{nlm2021medline}. Unlike these systems, which benefit from large collections of manually indexed training data and a domain-specific vocabulary, MATCH addresses multidisciplinary indexing without existing article-level annotations for its Concepts.

In the setting of MATCH, the challenge is to assign multiple relevant Concepts to each article from a large hierarchical taxonomy spanning diverse scientific disciplines, without existing article-level annotations. The system must identify a small number of suitable Concepts among many possible candidates while maintaining consistency with the taxonomy.

\subsection{Hierarchical Multi-Label Classification}
\label{subsec:xmlc}
Scientific document classification becomes more complex when several labels can be assigned to the same article and these descriptors are organized within a taxonomy.
In this setting, the possible assignments are not independent, because a specific topic is connected to broader parent classes.
This problem has been studied using the ACM Computing Classification System \citep{acm2012ccs}, a domain-specific framework for computer science.
Although this work provides an important example of hierarchical multi-label classification, it is limited to computer science and uses a smaller category space \citep{sadat2022}.

Another challenge concerns the size of the label space.
In supervised multi-label classification, only a small fraction of the available labels is typically relevant to each document, and many labels have few annotated examples.
This imbalance makes supervised learning more difficult, especially for specialized classes.
Research on large-scale multi-label prediction has examined this issue and shown that rare labels are difficult to learn when annotated examples are limited \citep{babbar2019}.

Extreme multi-label methods can handle label spaces of millions of labels, for example through label trees and attention-based models \citep{you2019attentionxml}, but they typically rely on labeled training data, which are not available for most of the Concepts considered here.
Recent work has also combined a bi-encoder, which retrieves candidate labels, with a large language model that selects among them in a zero-shot setting, for hierarchical multi-label classification of scientific documents \citep{tabatabaei2025llm}; the taxonomy considered there has several thousand nodes, substantially fewer than the one addressed by MATCH.

A related approach is retrieve-then-rerank, in which a fast first stage retrieves candidate labels and a more expensive model selects or ranks them.
This general pattern is well established in information retrieval \citep{nogueira2019passage}, and large language models have also been explored as zero-shot rerankers \citep{sun2023chatgpt}.

\subsection{From SCALE to Concept Assignment}
\label{subsec:from-scale}
The construction of a fine-grained scientific taxonomy and its application to individual documents are two distinct problems.
SCALE primarily addressed the first by transforming author keywords into Scientific Concepts and organizing them within the OpenAlex hierarchy \citep{raimondi2026scale}.
SCALE also identified article tagging as a possible application, but it did not develop a method for assigning Concepts to articles.

Building on SCALE, we address the next step: identifying which Concepts in the taxonomy best represent the scientific content of each article.
Relevant evidence may appear in the title, abstract, or keywords, while similar expressions can refer to different subjects across disciplines.
A practical solution must therefore identify a small number of suitable labels from a large candidate space without excluding important alternatives too early.

The preceding research highlights several challenges: existing publication-level classifications in scientometrics do not directly address assignment to the SCALE taxonomy, established indexing systems such as MeSH are domain-specific, and supervised extreme multi-label methods depend on annotated training data, while recent zero-shot approaches have been evaluated on smaller taxonomies.
MATCH addresses these challenges by combining retrieval, taxonomy-aware filtering, and language-model ranking in a scalable, production-deployed framework that assigns fine-grained Scientific Concepts to millions of publications and supports updates as new articles become available.

\section{Methods}
\label{sec:methods}

This section presents MATCH, the framework developed to assign SCALE
Concepts (Section~\ref{subsec:task-data}) to scientific articles. Given the title, abstract, and author keywords for each article, MATCH assigns a small ranked set of Concepts, each retaining its complete path through the taxonomy. The
main difficulty lies in selecting a coherent set of Concepts from a label
space of approximately 114{,}000 entries. Evidence for a Concept may be
concentrated in a single author keyword, a phrase in the title, or one
sentence of the abstract. Conversely, similar terminology may denote
different Concepts depending on the disciplinary context in which it
occurs.

This setting requires progressively reducing the search space while
preserving relevant alternatives. MATCH therefore separates the assignment
task into three stages with distinct roles. First, a multi-view retrieval
stage identifies a broad set of potentially relevant Concepts, prioritizing
coverage across different parts of the article. Second, deterministic
hierarchy-aware filters reduce this candidate set by enforcing disciplinary
consistency and removing redundant assignments. Third, a language-model
reranker resolves the remaining semantic ambiguity by selecting and ranking
the Concepts that best describe the article. This separation allows each
stage to address a different source of difficulty and limits the use of
language-model inference to a small candidate set.

\begin{figure}[H]
\centering
\includegraphics[width=\linewidth]{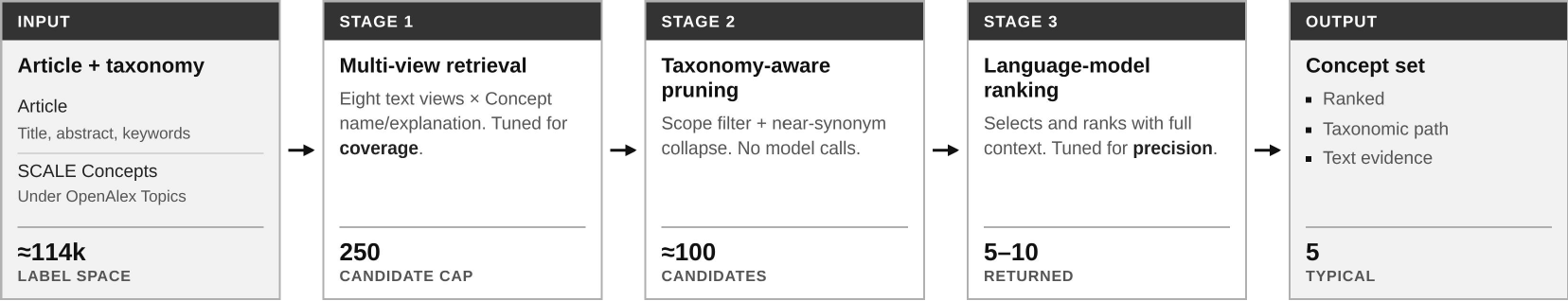}
\caption{Overview of the MATCH framework. Candidate Concepts are drawn from
the approximately 114{,}000 Concepts of the SCALE taxonomy and retrieved from
multiple views of the article (title, abstract, and author keywords) in
Stage~1. Stage~2 prunes them with deterministic, hierarchy-aware filters.
In Stage~3, a language model selects and ranks a small ordered set of
Concepts (at most ten), each retaining its full taxonomic path.}
\label{fig:workflow}
\end{figure}

The following subsections describe the data (Section~\ref{subsec:task-data}), the three processing stages (Sections~\ref{subsec:retriever}--\ref{subsec:reranker}), the modular deployment of the reranker (Section~\ref{subsec:deployment}) and the expert evaluation of two reranker configurations (Section~\ref{subsec:eval-design}). Figure~\ref{fig:workflow} summarizes the complete
workflow.

\subsection{Data}
\label{subsec:task-data}

MATCH uses the SCALE taxonomy \citep{raimondi2026scale} as the controlled
vocabulary for tagging scientific articles, so that each article is
described by fine-grained descriptors that remain connected to a shared
disciplinary hierarchy. SCALE extends the four-level OpenAlex
classification \citep{priem2022openalex} with a fifth level of Scientific Concepts derived from the author keywords of MDPI articles.
The resulting hierarchy forms a pyramid, with 4 Domains at the top, followed by 26 Fields, 252 Subfields, and 4{,}516 Topics. At the base are 113{,}892 Concepts, assigned to 4{,}461 of these Topics. Each Concept inherits a complete hierarchical path through its Topic, Subfield, Field, and Domain.

Each Concept has a unique identifier, a canonical name, and a short LLM-generated explanation of its meaning in the disciplinary context. Since Concepts may span multiple disciplines, the same canonical name can appear under different Topics as distinct entries, each with its own hierarchical path. For example, ``Industry 5.0'' appears under multiple Topics. MATCH therefore uses Concept identifiers as labels and leverages the hierarchy both to structure the output and to constrain candidate pruning (Section~\ref{subsec:pruning}).

MATCH was developed and demonstrated on the MDPI journal corpus, comprising
over 2 million articles as of October 2026
(Section~\ref{subsec:corpus-stats}). For each article, it takes as input only
the title, abstract, and author keywords. This keeps the input consistently
available across the corpus and avoids the processing and access requirements
of full text. MATCH can be applied to other scholarly records with the same inputs, although its transferability beyond the MDPI corpus remains to be evaluated (Section~\ref{subsec:limitations}).

\subsection{Stage 1: Multi-view Concept Retrieval}
\label{subsec:retriever}

With 113{,}892 Concepts and over 2 million articles, scoring every article-Concept pair requires approximately $2\times10^{11}$ similarity evaluations. To make this computationally feasible, MATCH uses embedding-based retrieval, enabling efficient similarity computation through GPU matrix multiplication. Article and Concept texts are encoded separately (bi-encoder) using the SPECTER2 base model\footnote{\url{https://huggingface.co/allenai/specter2_base}}, a scientific document encoder pretrained with a citation-aware contrastive objective \citep{singh2023}. Concept embeddings are precomputed and reused across articles, occupying only a few hundred MB of GPU memory. Cosine similarities are then computed through matrix multiplication, enabling efficient and exhaustive scoring of all Concepts for each article.

Because evidence for a Concept is often localized within an article, a single embedding of the entire article may obscure such evidence. MATCH therefore uses eight article-Concept view pairs (Table~\ref{tab:retrieval-pairs}), dividing the title, abstract, and author keywords into shorter \emph{segments} and matching them to either Concept names or explanations. Short segments with only a few words are matched to Concept names, while longer passages are matched to explanations, capturing both terminological matches and broader topical similarity. This stage aims to produce a high-recall candidate set for subsequent refinement. When keywords are unavailable, retrieval proceeds using the title and abstract; the framework can also operate using only the title. Across the corpus, retrieval requires approximately $10^{13}$ segment-to-Concept similarity computations, which remain tractable on GPU.

To retrieve relevant Concepts efficiently, we apply view-specific cosine similarity thresholds and retain at most the top-$k$ Concepts exceeding each threshold. More specific signals use higher thresholds: author keywords require near-exact similarity to Concept names, whereas abstract sentences are matched more permissively to Concept explanations. Since a Concept may be retrieved from multiple article segments and view pairs, MATCH considers both the strongest similarity and repeated evidence. A strong match indicates relevance to a particular part of the article, while repeated matches across different segments provide additional support for the Concept. The retrieval score is aggregated in three steps. First, within each view pair, each Concept retains its highest similarity across segments, and only the top-$k$ Concepts are kept. Second, candidates from all eight view pairs are merged by Concept identifier, retaining the highest cosine similarity as $s_{\mathrm{best}}$, without additional normalization. Third, Concepts retrieved by multiple distinct segments receive a logarithmic boost:
$
s_{\mathrm{final}} = s_{\mathrm{best}}(1 + \alpha \log h),
$
where $h$ is the number of distinct segments retrieving the Concept and $\alpha=0.1$. A Concept retrieved only once ($h=1$) retains its original score, while repeated matches increase the score with diminishing returns. The top 250 Concepts ranked by $s_{\mathrm{final}}$ are passed to Stage~2 for further refinement. The similarity thresholds, boost parameter $\alpha$, and candidate cap were selected heuristically during development.

Since Stage~1 prioritizes recall, these settings are deliberately permissive: irrelevant candidates can be removed in subsequent stages, whereas omitted Concepts cannot be recovered. The first column of Figure~\ref{fig:retrieval-evidence} illustrates the retrieved candidates. In this example, the Concept \textit{Community Pharmacy} is retrieved from both the author keyword view and the abstract sentence-window view, with the corresponding evidence highlighted in the article text. Figure~\ref{fig:concept-map-funnel} further illustrates the distribution of retrieved Concepts across the Concept map.

\begin{table}[H]
\centering
\caption{Article-Concept view pairs used in Stage~1, with per-pair
candidate limits ($k$) and minimum cosine similarity thresholds.
Title views include the complete title, three contiguous word groups,
and sliding three-word phrases. Abstract views include individual
sentences, overlapping three-sentence windows, an adaptive split
into three to six segments, and word-level fragments of each
sentence. Author keywords are processed individually.}
\label{tab:retrieval-pairs}
\begin{tabular}{llrr}
\toprule
Article view & Concept view & $k$ & Threshold \\
\midrule
Title, complete            & Explanation & 30 & 0.90 \\
Title, word groups         & Name        & 30 & 0.90 \\
Title, sliding n-grams     & Name        & 30 & 0.90 \\
Abstract, single sentences & Explanation & 50 & 0.87 \\
Abstract, sentence windows & Explanation & 50 & 0.87 \\
Abstract, adaptive split   & Explanation & 50 & 0.87 \\
Abstract, word segments    & Name        & 50 & 0.91 \\
Author keywords            & Name        & 30 & 0.94 \\
\bottomrule
\end{tabular}
\end{table}

\subsection{Stage 2: Taxonomy-aware Candidate Pruning}
\label{subsec:pruning}

Because Stage~1 favors coverage, its candidates include Concepts that are
close in embedding space but unsuitable as descriptors: they belong to an
unrelated Field, or they duplicate one another under different Topics.
Stage~2 removes them with two deterministic filters that use the SCALE
hierarchy and do not require large language model calls. This also reduces the input length and computational cost of Stage~3, an important consideration for corpus-scale deployment (Section~\ref{subsec:deployment}).

The retrieval stage may introduce Concepts from less relevant disciplines due to semantic similarities in the text. To reduce such noise, the first filter removes candidates outside the article's main disciplinary scope. It applies only to single-hit Concepts ($h=1$), while high-hit Concepts ($h>1$) are preserved because matches from multiple segments provide stronger evidence of relevance. Single-hit candidates must pass two tests. First, Fields are ranked by their frequency among candidates, and those in the rare tail, accounting for up to 15\% of candidates, are excluded. This reduces occasional cross-field matches that may be less relevant to the article. Second, Subfields are ranked by two measures: the number of candidates they contain and the proportion of their Concepts represented among the candidates. A Subfield is retained if it ranks within the top 60\% by either measure. The first measure favors Subfields with more matching Concepts, while the second preserves smaller Subfields with strong representation. The final set includes all high-hit Concepts and single-hit Concepts that pass both tests.

The retrieval stage may also return multiple candidates representing the same or closely related Concepts, leading to redundant assignments. This occurs because the same canonical name can appear under different Topics (Section~\ref{subsec:task-data}), or similar Concepts may have slightly different names. To reduce redundancy, the second filter groups duplicate and near-duplicate candidates and retains the most relevant representative from each group. For candidates sharing the same name, MATCH prefers those belonging to the article's dominant Fields, breaking ties by the token overlap between the Topic name and the article's title and keywords. Near-duplicate names are then grouped using a Jaccard similarity threshold of 0.50 on name tokens, supplemented by a character-level check for morphological variants such as singular and plural forms. Connected candidates are grouped using union-find. Candidates matching different author keywords are kept separate to preserve distinctions made by the authors. Each group retains one representative, prioritized by exact keyword match, keyword coverage of the Concept name, Topic-name overlap with the article context, and retrieval score. This filtering step reduces redundant candidates while preserving distinctions that may be important for accurately representing the article's scientific content.

The second column of Figure~\ref{fig:retrieval-evidence} shows an example of the remaining candidates after pruning, while Figure~\ref{fig:concept-map-funnel} illustrates the distribution of Concepts on the Concept map after the disciplinary scope filter and redundancy collapse.

\subsection{Stage 3: Language Model Selection and Ranking}
\label{subsec:reranker}

The pruned candidate set is still too large to serve as a descriptor list, and semantic similarity alone does not necessarily reflect a Concept's relevance to the article's scientific contribution. A Concept may share terminology with the article but describe only a peripheral aspect, or it may belong to a disciplinary path that does not fit the article. Stage~3 therefore uses a language model to select and rank the Concepts that best represent the article's scientific content. The reranker is model-agnostic: it receives the article title, abstract, and keywords together with the candidate Concepts, and its output is restricted to those candidates. The deployment configuration of the reranker is described in Section~\ref{subsec:deployment}.

The model performs four tasks that require reading the article rather than
comparing it with each Concept in isolation.
It (i) infers the article's primary discipline; (ii) discards
candidates that are generic to scientific writing, peripheral, or placed
on an unrelated disciplinary path when a discipline-appropriate
alternative exists; (iii) resolves remaining same-name or near-duplicate
candidates by preferring the path consistent with the article's
discipline; and (iv) ranks the surviving Concepts by centrality, from the core
contribution, through the primary subject and any Concept named in the title,
to supporting methodology; the rank of a Concept is its position in the
returned list. In multi-step model configurations, these
judgments are distributed across successive stages
(Section~\ref{subsec:deployment}).

\begin{figure}[H]
    \centering
    \includegraphics[width=\linewidth]{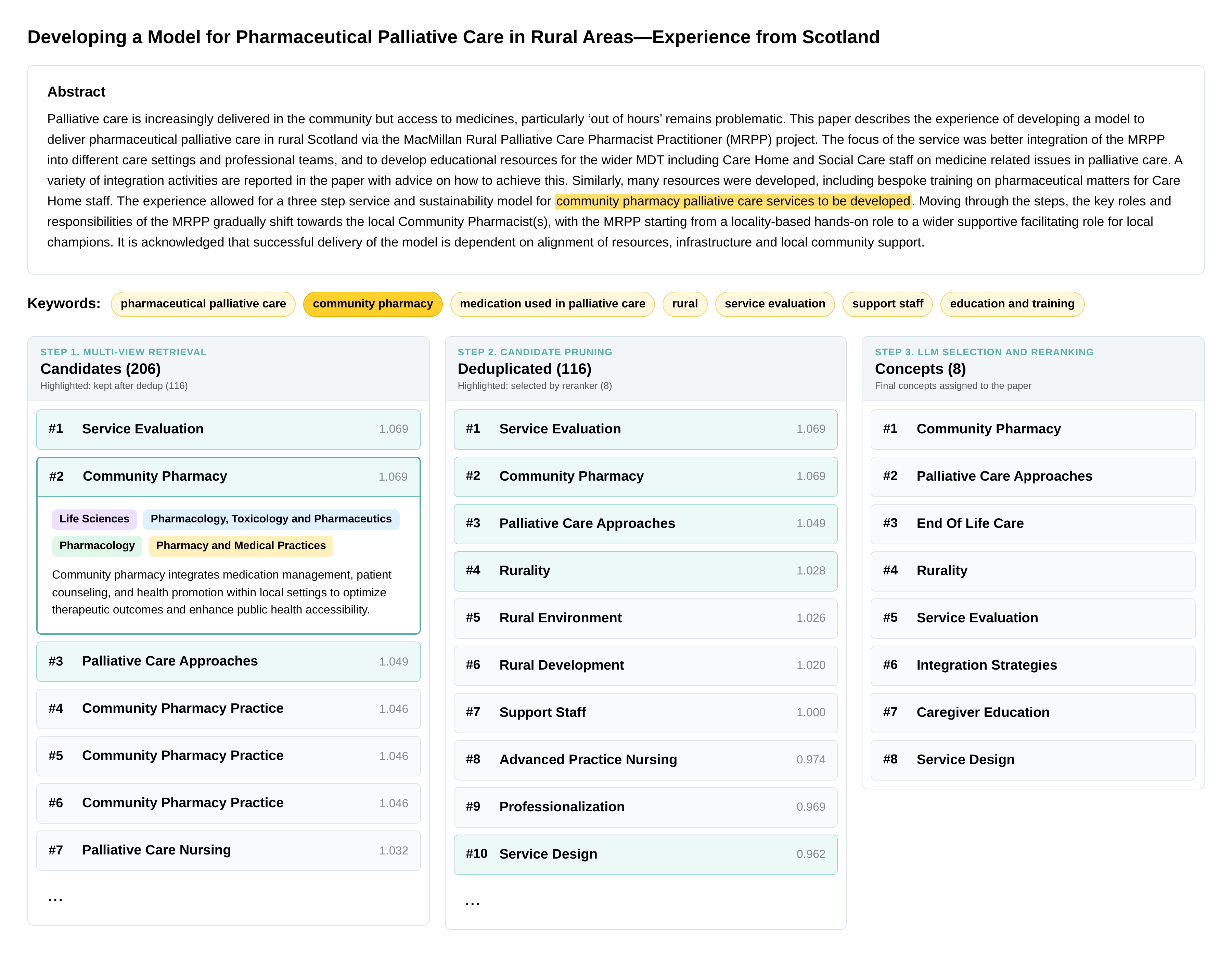}
    \caption{Retrieval evidence and Concept assignment for an illustrative article \citep{akram2017pharmaceutical}. The three panels show the MATCH pipeline: (1) multi-view retrieval of 206 candidate Concepts (Section~\ref{subsec:retriever}); (2) candidate pruning to 116 unique Concepts (Section~\ref{subsec:pruning}); and (3) LLM-based selection and reranking of eight final Concepts, listed in rank order (Section~\ref{subsec:reranker}). The values shown in the first two panels are boosted retrieval scores ($s_{\mathrm{final}}$; Section~\ref{subsec:retriever}). \textit{Community Pharmacy} is selected as an example to illustrate the retrieval evidence, with highlighted spans indicating the matching author keyword and abstract phrase, alongside its SCALE taxonomic path (Domain, Field, Subfield, Topic) and explanation. The same information is available for all retrieved Concepts but is shown only for \textit{Community Pharmacy} for clarity.}
    \label{fig:retrieval-evidence}
\end{figure}

\begin{figure}[H]
    \centering
    \includegraphics[width=0.95\textwidth,keepaspectratio]{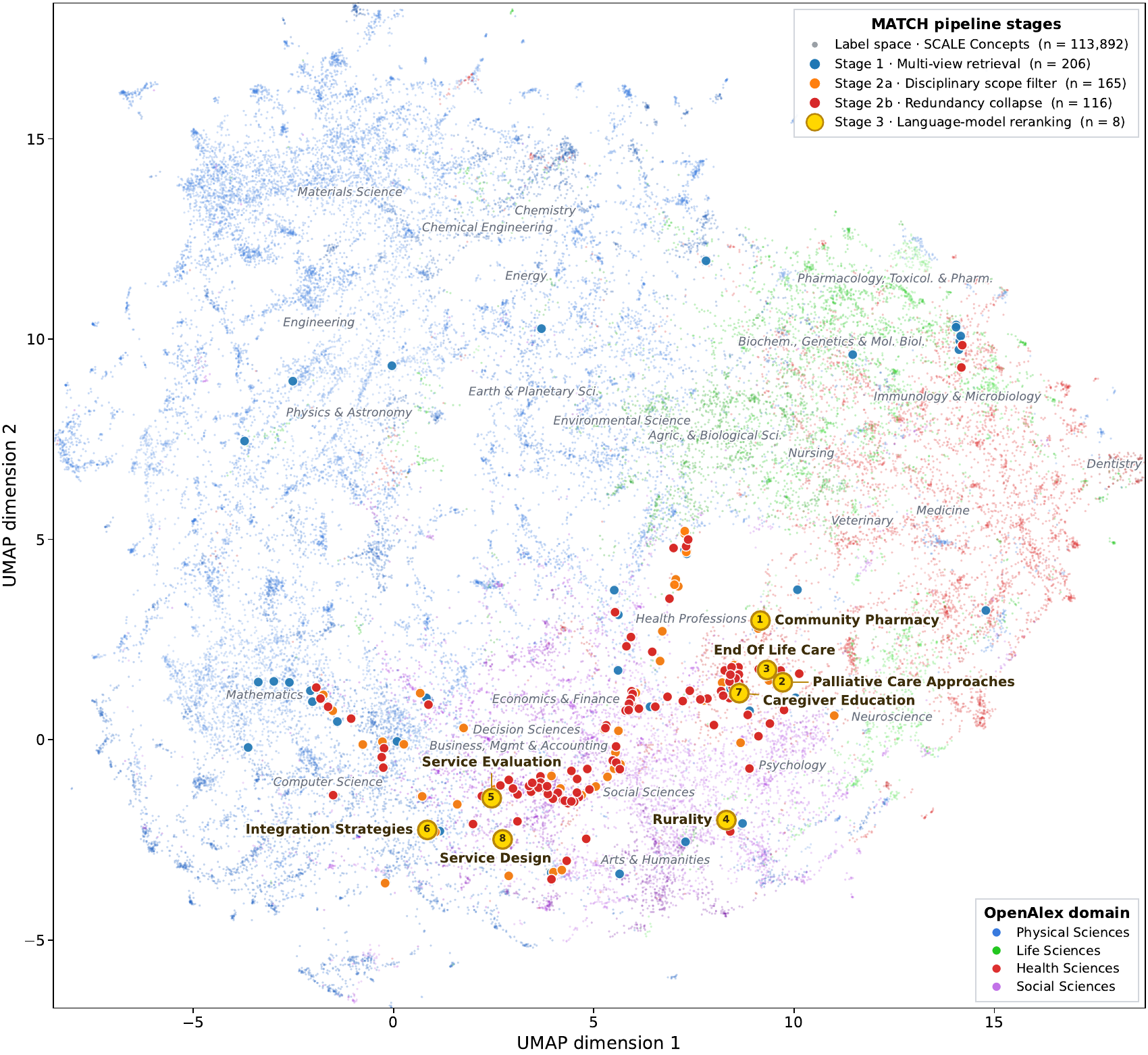}
    \caption{MATCH candidates on the SCALE Concept map for the article of
    Figure~\ref{fig:retrieval-evidence}, tagged with the production reranker (Configuration C1 in Section~\ref{subsec:eval-design}).
    Background points are the full
    Concept layer ($n=113{,}892$), colored by OpenAlex Domain; italic labels
    mark the 26 OpenAlex Fields. Larger markers show the article's candidates, colored according to their
    progress through the pipeline: retrieved at Stage~1 but removed by the
    disciplinary scope filter at Stage~2a; retained after Stage~2a but removed
    by redundancy collapse at Stage~2b; retained after Stage~2b but not selected
    by the language model at Stage~3; and selected as final Concepts at Stage~3
    (labeled). Numbers in
    the final markers give the rank assigned by the language model. Legend
    counts are candidate-set sizes after each stage. Coordinates are a 2D UMAP
    projection of SPECTER2 embeddings of each Concept's name concatenated with
    its explanation; map position reflects textual similarity, not
    taxonomic placement.}
    \label{fig:concept-map-funnel}
\end{figure}

The third column of Figure~\ref{fig:retrieval-evidence} shows an example of the final Concepts after language-model selection and ranking, while Figure~\ref{fig:concept-map-funnel} illustrates their names and locations on the Concept map.


\subsection{Modular Reranking and Deployment}
\label{subsec:deployment}

At corpus scale, the cost of Stage~3 (Section~\ref{subsec:reranker}) becomes a design constraint. With an API-based model, run by a provider and accessed through a paid API, cost grows linearly with the number of articles and recurs at every re-tagging, so per-token pricing can become prohibitive at corpus scale. MATCH therefore requires that the
reranker support both API-based and self-hosted inference, the latter
replacing per-token costs with the cost of the in-house GPU infrastructure.

\begin{figure}[H]
    \centering
    \includegraphics[width=\textwidth]{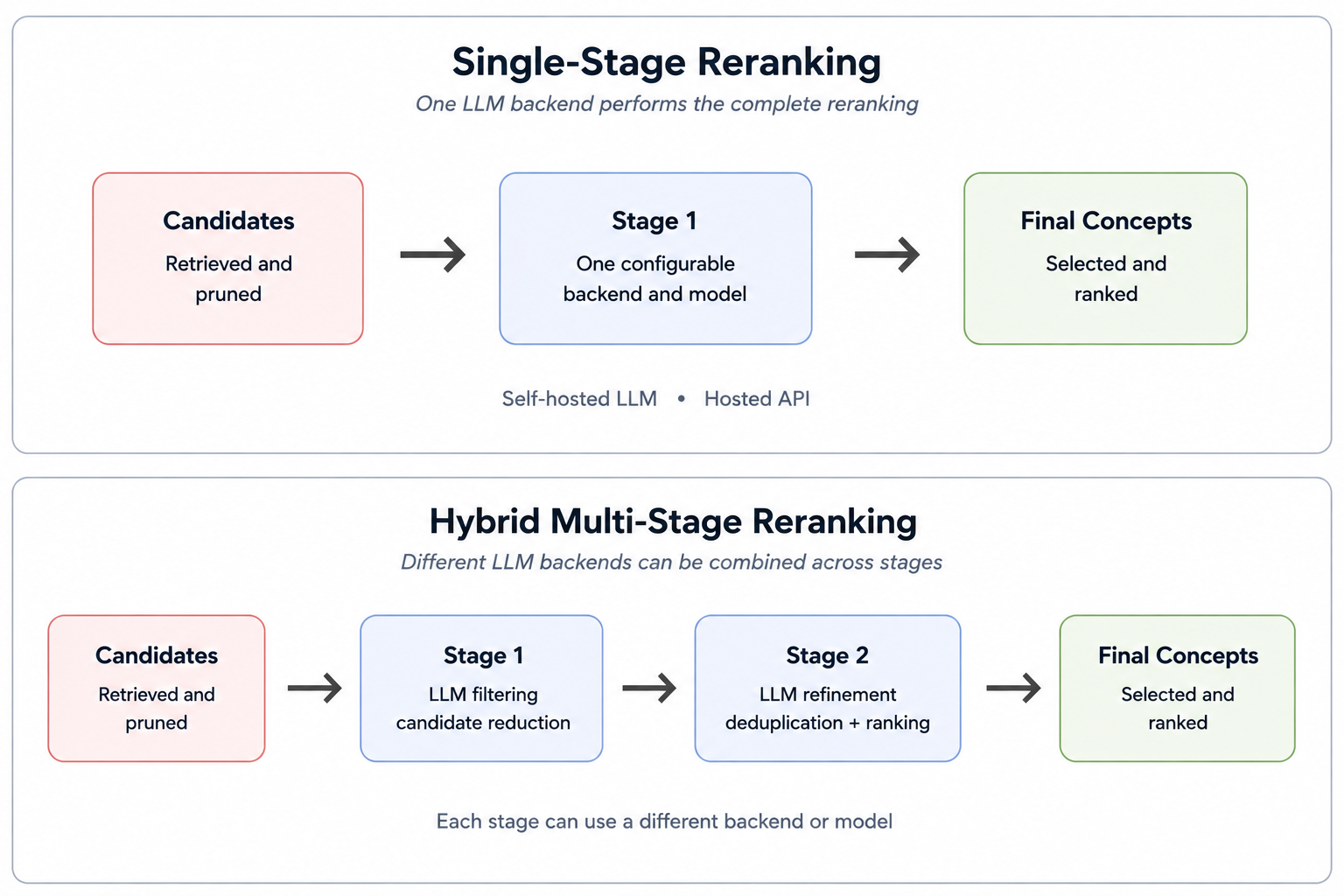}
    \caption{Configurable reranking deployments. MATCH supports both
    single-step reranking with a single LLM backend and multi-step
    configurations in which different steps may use different inference
    backends or models.}
    \label{fig:deployment}
\end{figure}

The reranker is implemented as a configurable sequence of steps rather than a single fixed model call (Figure~\ref{fig:deployment}). Each step specifies an inference backend, a model, and a maximum number of returned Concepts. The first step receives the Stage~2 candidates, while subsequent steps operate on the output of the preceding step, progressively reducing the candidate set. The number and purpose of these steps can be adapted to the capabilities of the selected model. In the API-based configuration, a hosted model performs the complete selection and ranking task described in Section~\ref{subsec:reranker}. In the self-hosted configuration, the task is divided into simpler steps to accommodate a model with more limited capabilities: intermediate steps remove peripheral, inconsistent, or duplicate candidates, and the final step ranks the remaining Concepts by their relevance to the article. This modular design allows models and processing steps to be replaced or adjusted without modifying Stages~1 and~2.

To ensure consistent and reliable assignments, structured output is enforced at every step through schema-constrained parsing for API-based models and constrained JSON decoding for self-hosted models. Returned Concepts must belong to the candidate set, preserve their original names, and appear only once, subject to each step's maximum output length. These constraints ensure that the model selects Concepts from the SCALE vocabulary rather than generating new labels. Invalid outputs are retried up to three times before a failure is reported. Each assignment also retains its Stage~1 retrieval score and supporting text segments, allowing assigned Concepts to be traced back to the article text (Figure~\ref{fig:retrieval-evidence}). Completed assignments are checkpointed, enabling interrupted processing to resume without repeating completed inference. Beyond the initial processing of historical publications, MATCH is deployed in production with a daily update pipeline that assigns Concepts to newly available articles. This enables continuous indexing as the literature grows, without requiring the entire corpus to be reprocessed.

\subsection{Expert Evaluation}
\label{subsec:eval-design}
\label{subsec:metrics}

To assess the flexibility of MATCH across different deployment configurations, we compare two operational rerankers: a self-hosted configuration (C1) and an API-based configuration (C2). C2 uses GPT-5.4 mini~\citep{gpt54mini}, accessed through Azure OpenAI, as a single-step reranker chosen to balance quality and cost. C1 uses a three-step chain based on the self-hosted Qwen3-VL-32B-Instruct-FP8 model (hereafter Qwen3 32B~\citep{bai2025qwen3vl}), with two filtering steps followed by one ranking step. This decomposition reduces the complexity of each inference step and aims to achieve assignment quality comparable to the API-based configuration. Both configurations operate on the same Stage~2 candidate sets and return at most ten Concepts. Because the configurations differ in both model and chain design, the comparison evaluates the complete reranking systems rather than isolating the effects of model choice, multi-step processing, or self-hosting.

The relevance of the returned Concepts was assessed by domain experts to cover the disciplinary diversity of the corpus. This expert study was previously summarized in SCALE \citep{raimondi2026scale}; here, we report the full evaluation protocol, comparison of C1 and C2, and reliability of the judgments. The evaluation focuses on the relevance of assigned Concepts rather than the structure of the SCALE taxonomy. Twenty-two MDPI employees holding a doctorate participated as experts, two for each of 11 scientific fields. For each field, 10 articles were randomly sampled from MDPI publications and independently assessed by both experts, yielding 110 unique articles and 220 planned article--evaluator assessments.  Each expert was shown the article's title, abstract, and keywords alongside the Concept lists returned by C1 and C2, each containing at most ten Concepts. The lists were presented side by side without revealing their configurations, and their left/right positions were randomized for each article. Experts provided a binary relevance judgment for each Concept, with its explanation and taxonomic path available on demand to clarify ambiguous names. They also listed potential missing Concepts, which were used only to compute Recall. A Concept was considered relevant if it covered a topic of the article, even if its wording differed from what the expert would have chosen. For example, acronyms expanded to their full forms, compound keywords represented by two Concepts, and verbose phrases reduced to their core terms were considered correct.

Evaluation metrics are computed separately for each article--evaluator pair and then averaged. Let $C$ denote the ranked list of Concepts returned by a configuration, $C_k$ its first $k$ entries, $R \subseteq C$ the Concepts judged relevant, and $m$ the number of missing Concepts reported by the expert. Precision and Recall are defined as
\begin{equation}
\label{eq:precision-recall}
\mathrm{Precision} = \frac{|R|}{|C|},
\qquad
\mathrm{Recall} = \frac{|R|}{|R| + m}.
\end{equation}
F1 is the harmonic mean of Precision and Recall. Since F1 is calculated for each article--evaluator pair before averaging, the mean F1 generally differs from the harmonic mean of the mean Precision and Recall.

To assess ranking quality, we use normalized discounted cumulative gain (NDCG) \citep{jarvelin2002cumulated}. Let $r_i \in \{0,1\}$ indicate whether the Concept at rank $i$ is relevant. Then,
\begin{equation}
\label{eq:ndcg}
\mathrm{NDCG@}k =
\frac{\sum_{i=1}^{k} r_i / \log_2(i+1)}
{\sum_{i=1}^{k} r_i^{*} / \log_2(i+1)},
\end{equation}
where $r_i^{*}$ denotes the relevance judgments reordered with relevant Concepts first. NDCG over the full list uses $k = |C|$ and measures how well relevant Concepts are ranked among those returned, without accounting for missing Concepts. We also compute Precision at $k$ \citep{manning2008introduction}:
\begin{equation}
\label{eq:precision-k}
\mathrm{P@}k = \frac{|C_k \cap R|}{|C_k|}.
\end{equation}
When fewer than $k$ Concepts are returned, both P@$k$ and NDCG@$k$ are computed using the actual list length. Reported scores are unweighted arithmetic means across article--evaluator pairs.

We compare C1 and C2 using paired Wilcoxon signed-rank tests \citep{wilcoxon1945individual} on article-level scores. Inter-rater agreement is evaluated on Concepts assessed by both experts, separately for each field and across all fields. Given the high prevalence of positive relevance judgments, we report observed agreement $p_o$, prevalence-adjusted bias-adjusted kappa (PABAK) \citep{byrt1993bias}, and Gwet's AC1 \citep{gwet2008computing}.

\section{Results}
\label{sec:results}
This section presents the empirical evaluation of MATCH from two complementary perspectives. Section~\ref{subsec:corpus-stats} reports the production-scale assignment of SCALE Concepts across the full MDPI corpus. Section~\ref{subsec:expert-eval} assesses the relevance of the assigned Concepts through a domain-expert evaluation, and Section~\ref{subsec:reliability} examines the reliability and consistency of the resulting expert judgments. As described in Section~\ref{subsec:eval-design}, the expert evaluation compares two configurations, C1 and C2.

\subsection{Large-scale Assignment and Taxonomy Coverage}
\label{subsec:corpus-stats}

MATCH was applied to the full MDPI corpus described in
Section~\ref{subsec:task-data}, comprising 2{,}017{,}235 articles, with the self-hosted configuration C1 (Section~\ref{subsec:eval-design}). The initial indexing of the historical corpus was completed over approximately two and a half months, with processing running for an average of 14 hours per day, corresponding to an estimated throughput of 1{,}900 articles per hour. MATCH has since been integrated into a daily production pipeline, processing approximately 900 newly published articles per weekday and 400 per day on weekends. These results demonstrate the practical feasibility of MATCH for both large-scale historical indexing and continuous indexing of new publications using self-hosted LLM inference.

The resulting assignments covered 94\% of the 113{,}892 Concepts in the taxonomy, indicating that the assigned Concepts were not confined to a small, frequently occurring subset. Concept coverage is defined as the proportion of taxonomy Concepts assigned to at least one article in the indexed corpus. The median number of Concepts assigned per article was five, indicating that MATCH typically represents an article using a compact set of Concepts while allowing multiple aspects of its content to be captured.

\begin{figure}[H]
    \centering
    \includegraphics[
        width=0.95\textwidth,
        keepaspectratio
    ]{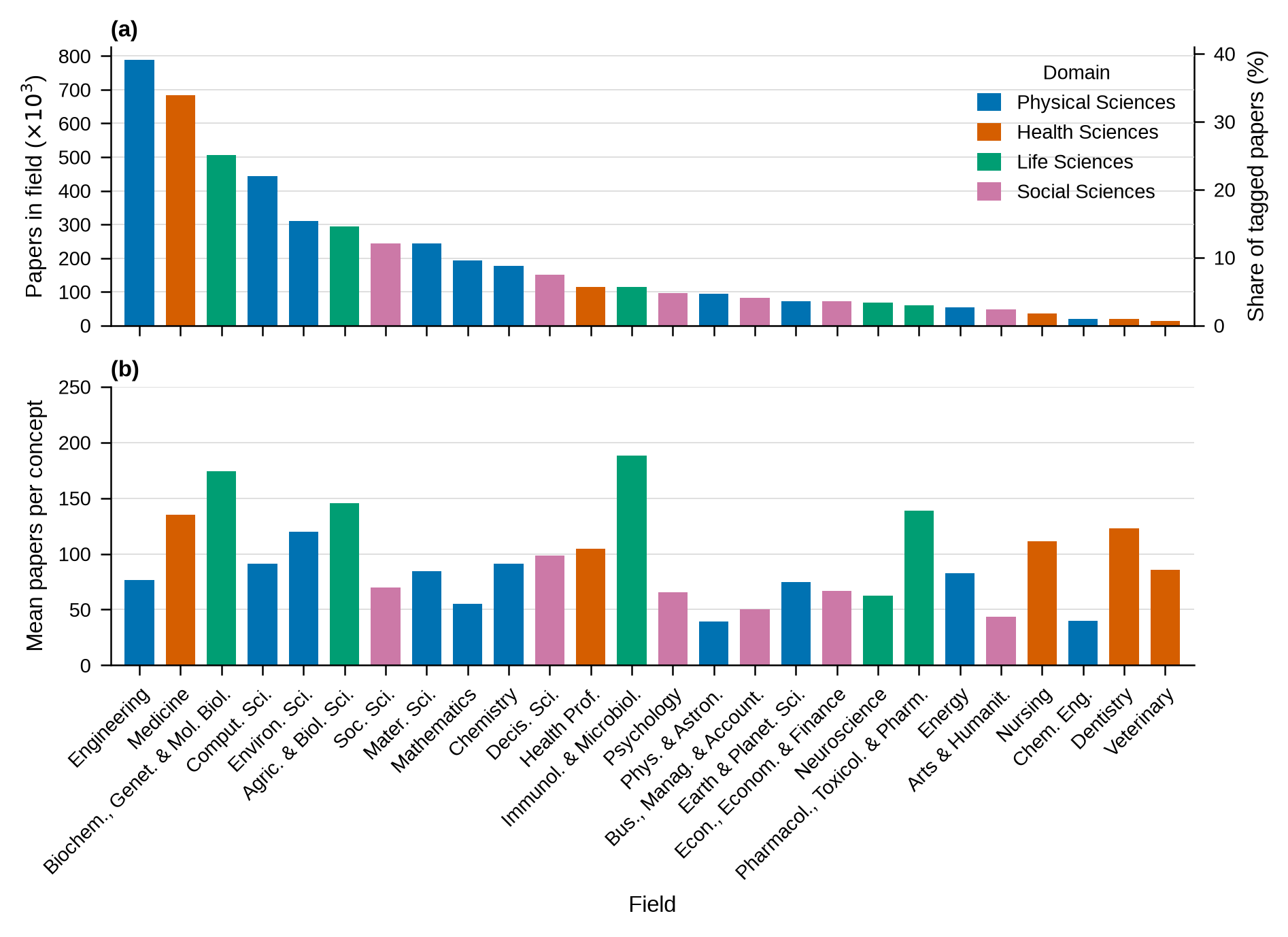}
    \caption{Distribution of article and Concept assignments across
    scientific fields. (a) Number of articles assigned at least one
    Concept in each field (left axis) and the corresponding percentage
    of all tagged articles (right axis). Because an article may be
    associated with multiple fields, field-level percentages are not
    mutually exclusive and therefore do not sum to 100\%.
    (b) Mean number of articles per Concept within each field, calculated
    over Concepts observed at least once in the corpus. Fields are ordered
    by the number of associated articles in panel~(a), and colors denote
    broad scientific domains.}
    \label{fig:distribution}
\end{figure}

Figure~\ref{fig:distribution} summarizes the distribution of Concept
assignments across scientific fields. Figure~\ref{fig:distribution}(a)
reports the number and proportion of articles associated with each field.
Article counts vary substantially across fields, potentially reflecting the disciplinary composition of the MDPI corpus rather than bias in MATCH. Figure~\ref{fig:distribution}(b) reports the mean number of articles per observed Concept within each field. Some fields with relatively few associated articles exhibit comparatively high mean numbers of articles per Concept. 
Thus, field-level article volume does not correspond directly to the average frequency with which individual Concepts are assigned.
Concept frequencies are not expected to be uniform within fields, as their uneven distribution may reflect variation in the prevalence of research topics across the MDPI corpus.

As individual articles may be assigned multiple Concepts, and these Concepts may map to different fields and domains, the assignments also provide an indicator of cross-field topical breadth. Table~\ref{tab:interdisciplinary_papers} presents the distribution of articles according to the number of distinct fields represented by their assigned Concepts. Most articles were associated with either two fields (833{,}171; 41.30\%) or three fields (678{,}505; 33.64\%). A further 11.36\% were associated with four fields and 1.28\% with five or more, whereas 12.42\% were associated with only one field. Overall, 87.58\% of articles received Concepts spanning at least two fields. These findings demonstrate that MATCH can represent articles using Concepts from multiple scientific fields, rather than restricting each article to a single-field classification. However, multi-field assignments do not necessarily indicate interdisciplinary research; they only show that an article contains Concepts associated with different fields.

\begin{table}[!htbp]
\centering
\caption{Distribution of articles by the number of distinct fields
represented by their assigned Concepts. An article is associated with a
field when at least one assigned Concept maps to that field. Articles
associated with five or more fields are grouped in the final category.
Percentages are calculated over the full corpus
($N = 2{,}017{,}235$).}
\label{tab:interdisciplinary_papers}
\begin{tabular}{rrr}
\toprule
Number of fields & Articles & Articles (\%) \\
\midrule
1  & 250{,}567 & 12.42 \\
2  & 833{,}171 & 41.30 \\
3  & 678{,}505 & 33.64 \\
4  & 229{,}126 & 11.36 \\
5+ & 25{,}866  & 1.28 \\
\bottomrule
\end{tabular}
\end{table}
\FloatBarrier

\subsection{Expert Assessment of Concept Assignment}
\label{subsec:expert-eval}

Our expert assessment involved 22 experts across 11 scientific fields. Within each field, two experts independently evaluated the same 10 articles from their domain, yielding 110 unique articles and 219 article--evaluator assessments. One Physics article was assessed by only one expert because the other considered it outside their area of expertise. Evaluations were conducted independently without communication between paired experts. As four experts were from Materials Science, they were divided into two pairs (Materials Science 1 and Materials Science 2), each evaluating a different set of articles.

We evaluate Concept assignment quality using Precision, Recall, F1, and NDCG \citep{manning2008introduction,jarvelin2002cumulated}. To specifically assess the quality of the highest-ranked assignments, we additionally report Precision@1, Precision@5, and NDCG@5, with $k=5$ chosen to reflect the median of five Concepts assigned per article in production (Section~\ref{subsec:corpus-stats}). As shown in Table~\ref{tab:eval-overall}, the two configurations achieve comparable performance, with C2 showing slightly higher overall Precision and NDCG, while C1 achieves slightly higher Precision@1 and NDCG@5. Recall and F1 are also reported, but should be interpreted cautiously because missing Concepts were identified by experts rather than through exhaustive annotation of the taxonomy. Consequently, these metrics provide estimates based on expert-reported omissions rather than exhaustive ground truth.

For the primary inferential analysis, evaluator scores were averaged within each article, and C1 and C2 were compared using paired Wilcoxon signed-rank tests over the resulting 110 article-level pairs. No statistically significant differences were found in overall Precision ($p=0.11$) or NDCG ($p=0.79$). A sensitivity analysis at the article--evaluator level ($n=219$) yielded $p=0.11$ and $p=0.58$, respectively, although assessments of the same article are not independent. Exploratory comparisons at the top-five cutoff likewise revealed no significant differences in Precision@5 ($p=0.99$) or NDCG@5 ($p=0.97$). Although these results do not establish statistical equivalence or rule out practically meaningful differences, the observed scores are descriptively similar. Together with the ability to avoid external LLM API charges, these findings support the selection of the self-hosted configuration for large-scale production indexing.

\begin{table}[!htbp]
\centering
\caption{Aggregate expert evaluation results for the two reranking configurations. Precision@1, Precision@5, and NDCG@5 evaluate the highest-ranked Concept assignments, whereas Precision, Recall, F1, and NDCG are computed over all assigned Concepts. Results are aggregated over 219 article--evaluator assessments,
corresponding to 110 unique articles.}
\label{tab:eval-overall}
\begin{tabular}{lrr}
\toprule
Measure (\%) & C1 (Qwen3 32B)  & C2 (GPT-5.4 mini) \\
\midrule
Precision@1  & 95.9 & 94.5 \\
Precision@5  & 88.7 & 88.9 \\
NDCG@5       & 97.4 & 96.9 \\
Precision & 81.5 & 83.3 \\
Recall    & 96.0 & 95.6 \\
F1         & 86.7 & 87.8 \\
NDCG & 96.7 & 97.0 \\ 
\bottomrule
\end{tabular}
\end{table}
\FloatBarrier

Performance varies across fields (Table~\ref{tab:eval-by-field}). Biochemistry and History achieve the highest Precision@5, whereas Physics, Mathematics, and Materials Science 1 show comparatively lower precision. Differences between the two configurations also vary by field; for example, C1 achieves higher Precision@5 in Physics, whereas C2 achieves higher Precision@5 in Molecular and Cellular Biology.

\begin{table}[!htbp]
\centering
\caption{Field-level Precision@5 for the two reranking configurations. Each field comprises 10 papers independently evaluated by two experts from the corresponding domain, except one Physics paper rated by a single expert. Cell values are unweighted arithmetic means over article--evaluator pairs, rounded to one decimal place; $n$ is the number of distinct papers.
The Overall row is the mean over all 219 pairs.}
\label{tab:eval-by-field}
\begin{tabular}{lrrr}
\toprule
Field & $n$ & C1 Precision@5 (\%) & C2 Precision@5 (\%) \\
\midrule
Physics                           & 10 & 85.0 & 75.0 \\
Materials Science 1               & 10 & 79.0 & 78.0 \\
Mathematics                       & 10 & 81.0 & 80.8 \\
Atmospheric/Environmental Science & 10 & 85.0 & 90.0 \\
Engineering                       & 10 & 88.0 & 90.0 \\
Botany/Plant science              & 10 & 88.0 & 91.0 \\
Drug Development and Delivery     & 10 & 95.0 & 91.0 \\
Molecular and Cellular Biology    & 10 & 87.0 & 93.0 \\
Materials Science 2               & 10 & 93.0 & 95.0 \\
Biochemistry                      & 10 & 98.0 & 97.0 \\
History                           & 10 & 97.0 & 97.0 \\
\midrule
Overall                           & 110 & 88.7 & 88.9 \\
\bottomrule
\end{tabular}
\end{table}
\FloatBarrier

\subsection{Reliability of Expert Judgments}
\label{subsec:reliability}

To assess the reliability of expert judgments, we measure inter-evaluator agreement on Concept relevance using observed agreement ($p_o$), prevalence-adjusted bias-adjusted kappa (PABAK), and Gwet's AC1 \citep{byrt1993bias,gwet2008computing}. As shown in Table~\ref{tab:eval-agreement}, overall observed agreement is 0.76, with a PABAK of 0.52 and an AC1 of 0.66. Agreement varies substantially across fields, with the highest values in Biochemistry (AC1 = 0.87) and History (AC1 = 0.84), and the lowest in Mathematics (AC1 = 0.22). This variation highlights differences in the consistency of expert relevance judgments across disciplines and should be considered when interpreting field-level Concept assignment precision.

Comparing Tables~\ref{tab:eval-by-field} and~\ref{tab:eval-agreement}, fields with higher inter-evaluator agreement also tend to exhibit higher Precision@5. Biochemistry and History, for example, show both high agreement and high precision, whereas Mathematics and Materials Science 1 show lower values on both measures. This suggests that lower precision in some fields may partly reflect greater disagreement among experts about Concept relevance, rather than lower tagging quality alone.

\begin{table}[!htbp]
\centering
\caption{Inter-evaluator agreement on Concept relevance by scientific field. Agreement is computed between the two experts within each field for Concepts assessed by both evaluators. We report the number of Concepts evaluated ($n$), observed agreement ($p_o$), prevalence-adjusted bias-adjusted kappa (PABAK), and Gwet's AC1.}
\label{tab:eval-agreement}
\begin{tabular}{lrrrrr}
\toprule
Field & $n$ & $p_o$ & PABAK & AC1 \\
\midrule
Biochemistry                      & 181 & 0.88 & 0.77 & 0.87 \\
History                           & 177 & 0.87 & 0.74 & 0.84 \\
Materials Science 2               & 182 & 0.83 & 0.66 & 0.79 \\
Molecular and Cellular Biology    & 185 & 0.83 & 0.65 & 0.78 \\
Drug Development and Delivery     & 178 & 0.80 & 0.60 & 0.75 \\
Botany/Plant science              & 179 & 0.77 & 0.54 & 0.70 \\
Engineering                       & 186 & 0.74 & 0.47 & 0.61 \\
Physics                           & 175 & 0.72 & 0.44 & 0.55 \\
Atmospheric/Environmental Science & 181 & 0.72 & 0.44 & 0.57 \\
Materials Science 1               & 181 & 0.66 & 0.33 & 0.37 \\
Mathematics                       & 174 & 0.55 & 0.09 & 0.22 \\
\midrule
Overall                           & 1979 & 0.76 & 0.52 & 0.66 \\
\bottomrule
\end{tabular}
\end{table}
\FloatBarrier

\begin{figure}[H]
\centering
\includegraphics[width=0.7\linewidth]{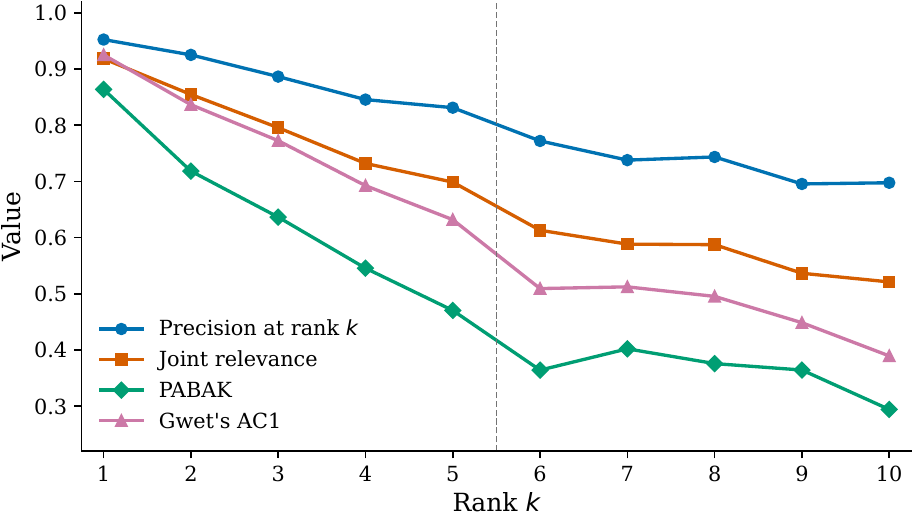}
\caption{Relevance and inter-evaluator agreement by rank, pooled across C1 and C2 (Section~\ref{subsec:eval-design}). Precision at rank $k$ is the fraction of Concepts at position $k$ judged relevant. ``Joint relevance'' is the fraction judged relevant by both experts evaluating the article. PABAK and Gwet's AC1 are computed separately at each rank, whereas Table~\ref{tab:eval-agreement} pools judgments across all ranks. The dashed line marks the boundary after the operating cutoff $k=5$.}
\label{fig:rank-marginal}
\end{figure}

Figure~\ref{fig:rank-marginal} shows how relevance and agreement vary by Concept rank. All measures generally decline across the returned list. Marginal precision decreases from approximately 0.95 at rank 1 to 0.83 at rank 5 and 0.70 at ranks 9--10. Joint relevance and agreement also decrease notably after rank 5. The top-five cutoff therefore captures the more reliable portion of the ranking and corresponds to the median number of Concepts assigned per article in production.

\section{Discussion}
\label{sec:discussion}
MATCH demonstrates that a fine-grained scientific taxonomy can be systematically applied to millions of publications, enabling Scientific Concepts to serve as structured descriptors of individual articles. The expert evaluation also supports the practical viability of a self-hosted reranking configuration as an alternative to API-based inference. We discuss the scientometric implications of Concept-level assignments, their relationship with SCALE, study limitations, and future directions.

\subsection{Concept-Level Assignments as a Unit of Analysis for Scientometrics}
\label{subsec:contribution}
The main contribution of MATCH is not limited to the assignment pipeline itself.
By linking individual publications to a small number of Scientific Concepts, the framework provides a standardized unit for describing scientific content with greater specificity than broad scholarly topics.
Previous work has shown that the granularity of publication-level classifications influences the type of bibliometric analysis that can be performed, and that topic-level systems can provide meaningful units for studying the literature \citep{waltman2012,sjogarde2018}.
MATCH extends this principle to a finer granularity, where multiple descriptors can represent different aspects of the same paper while remaining connected to an explicit taxonomic path.

A particularly relevant application of these assignments is the identification of new areas of study.
Existing scientometric approaches have detected emerging topics by identifying groups of publications that are both recent and rapidly growing \citep{small2014}.
Concept-level tagging offers a complementary basis for this purpose.
Because the same descriptor is used consistently across articles, its presence can be tracked over time and across scientific fields.
This makes it possible to study when an area appears, how quickly it grows, and how widely it spreads.
Developing such indicators is beyond the scope of the present study, but it represents a natural continuation of this work.

\subsection{Relation to SCALE and Study Limitations}
\label{subsec:limitations}
SCALE and MATCH should be interpreted as complementary components of the same framework.
SCALE establishes the controlled vocabulary and its hierarchical organization, whereas MATCH provides the mechanism for applying Scientific Concepts to individual publications at corpus scale \citep{raimondi2026scale}.
The expert evaluation measures whether the Concepts assigned by MATCH are relevant to each article.
At the same time, applying the vocabulary to over two million papers demonstrates that the Concept layer can be used operationally across a large and multidisciplinary collection.

An important property of MATCH is its modular design, which allows the language model used for final selection to be replaced without redesigning the overall framework. The evaluated self-hosted and API-based configurations achieved descriptively similar Precision and NDCG, with no statistically significant differences detected by paired tests. These findings support the practical viability of the evaluated self-hosted configuration as an alternative to API-based models, avoiding per-token API charges and offering a potential path toward more scalable deployment as model capabilities and computational requirements evolve.

Several limitations remain. First, the evaluation is restricted to MDPI publications. Because SCALE Concepts were derived partly from author keywords in this corpus, the results may not generalize to other collections. Second, the disciplinary scope filter in Stage~2 may exclude relevant Concepts from interdisciplinary articles when they are retrieved from only one text segment (Section~\ref{subsec:pruning}). Although the exemption for candidates retrieved from multiple segments mitigates this risk, it does not eliminate it. Third, the participating experts were MDPI employees. While the configurations were presented without identifying their source, familiarity with the corpus may influence absolute relevance judgments. Finally, Stage~3 relies on language-model inference, which may produce different Concept selections or rankings across repeated runs. Updates to hosted models introduce an additional source of variation, potentially affecting reproducibility.

\subsection{Outlook}
\label{subsec:outlook}
A next step is to extend the assignment of Scientific Concepts to the full Scilit corpus (\url{https://www.scilit.com/}).
This broader coverage would allow MATCH to be tested on a much larger and more diverse body of literature than the collection considered here.
In parallel, newer self-hosted models can be evaluated by domain experts using a procedure similar to that applied here to C1 and C2.

Concept-level tagging also provides a basis for studying how scientific activity changes over time.
Tracking the presence and growth of specific taxonomic entries could help identify emerging or declining research areas.
The same analysis could consider single Concepts or groups of related labels.
MATCH therefore provides a bridge from describing individual articles to examining how defined areas of science evolve over time.

Another direction is to examine how different research entities are connected.
The existing taxonomy mainly captures hierarchical links.
Future work could use associations observed across papers to build a network in which nodes correspond to Concepts and edges capture their connections.
These relationships could provide the basis for developing an ontology that combines the hierarchical organization of the taxonomy with the broader connections among research areas.
This could also support applications such as peer-reviewer recommendation and the identification of potential collaborators with related expertise.

\section{Conclusions}
\label{sec:conclusions}
MATCH enables the assignment of Scientific Concepts to individual publications within a taxonomy of approximately 114,000 fine-grained labels. Applied to over two million publications, with daily updates for newly available articles, MATCH demonstrates the feasibility of Concept-level assignment at corpus scale. Domain expert evaluation showed comparable observed Precision and NDCG across two reranking configurations, with no statistically significant differences detected, although equivalence was not established. Together with its production deployment, these results support the practical viability of the self-hosted configuration. By operationalizing the SCALE taxonomy at publication level, MATCH provides a foundation for fine-grained scientometric analyses, including the identification of emerging research areas and the study of scientific developments over time.
\section*{Acknowledgements}
The authors thank MDPI for providing the GPU infrastructure used in this
work. The authors also thank Dietrich Rordorf and Luke Sui for their
continuous support of this project. Finally, we thank the domain experts who
participated in the evaluation and generously contributed their time and
expertise.

\section*{Statements and Declarations}

\subsection*{Funding}
This research did not receive external funding.

\subsection*{Competing Interests}
All authors are employees of MDPI AG, the publisher of the journal corpus
used in this study. The study uses titles, abstracts, and author keywords of
MDPI publications. The authors declare no other financial or non-financial
interests that are directly or indirectly related to this work.
The domain experts who took part in the evaluation are also MDPI
employees.

\subsection*{Ethics Approval and Consent to Participate}
The expert evaluation involved adult professionals who voluntarily agreed to
take part and were informed of the purpose of the study. Participants judged
the relevance of Concepts assigned to published articles; no personal or
sensitive data about participants were collected or reported. Formal ethics
approval was therefore not sought.

\subsection*{Consent for Publication}
Not applicable.

\subsection*{Data Availability}
The MDPI articles used in this study are openly available on the MDPI
website. The SCALE taxonomy is described in \citet{raimondi2026scale}. Expert relevance judgments are available from the corresponding author on
reasonable request.

\subsection*{Code Availability}
The source code developed for this study is not publicly available because
it forms part of a proprietary internal MDPI research pipeline. Access to
relevant portions of the code may be considered upon reasonable request to
the corresponding author, subject to institutional approval and applicable
licensing restrictions.

\subsection*{Use of Generative AI and AI-Assisted Technologies}
Large language models are components of the method described in this study
(Section~\ref{subsec:deployment}). Separately, AI-assisted tools were used
for language editing, for drafting and revising passages of the Introduction
and Related Work, and to assist the verification of bibliographic details.
All AI-assisted output was reviewed and revised by the authors, every
reference was verified against its source, and the authors assume full
responsibility for the content of the manuscript.

\subsection*{Author Contributions}
Daniele Raimondi: Methodology, statistical analysis, coding, and manuscript
preparation.

Feichi Lu: Methodology, statistical analysis, coding, validation, and
manuscript revision.

Oliver Grun: Coding, analysis, and validation.

Mariia Eremina: Coding and validation.

Andrea Perlato: Scientific direction, supervision, manuscript preparation,
and manuscript revision.

All authors read and approved the final manuscript.

\bibliographystyle{unsrtnat}
\bibliography{references}

\end{document}